\documentclass[a4paper,11pt]{article}
\usepackage{jinstpub} % for details on the use of the package, please see the JINST-author-manual
\usepackage{lineno}
\usepackage{comment}

\newcommand{\lphi}{$\tau^- \rightarrow \ell^-\phi$}

\newcommand {\emep} {$e^{+}e^{-}$}
\newcommand {\deltae} {$\Delta E_{\tau}$}
\newcommand {\invmass} {$M_{\tau} $}

\newcommand {\invfem}{~fb$^{-1}$}

\newcommand{\gevcc}{GeV/$c^2$}
\newcommand{\mevcc}{MeV/$c^2$}

\newcommand {\taupair} {$e^{+}e^{-}\rightarrow \tau^+ \tau^-$}
\newcommand {\qqbar} {$e^{+}e^{-}\rightarrow q\bar{q}$}
\newcommand {\epem} {$e^{+}e^{-}$}
\newcommand {\myprocess} {$e^{+}e^{-}\rightarrow \mu^+\mu^-Z'$}

\newcommand {\taulalpha} {$e^{+}e^{-}\rightarrow \tau^+(\to \ell \alpha) \tau^- (\to \pi^+\pi^-\pi^-\nu)$}

\newcommand {\cmenergy} {$\sqrt{s}= 10.58$~GeV}

\title{\boldmath Studies on $\tau$ decays at Belle II}

\author{L. Zani}
\affiliation{INFN of Roma Tre,\\
Via della Vasca Navale, 84 Roma, Italy}

\emailAdd{laura.zani@roma3.infn.it}

\abstract{ Tau leptons %offer a clean environment to study the process of hadronization and 
are powerful tools to probe physics beyond the Standard Model (SM). % via precision mesurement of their properties or as direct searches for forbidden processes involving tau decays
The Belle II experiment is installed at the SuperKEKB asymmetric energy electron-positron collider and aims at collecting the world's largest sample of tau pair events \taupair. 
Direct searches for new invisible mediators, charged lepton flavor violation in $\tau$ decays, and tests of the SM via precision measurements of $\tau$ lepton properties and couplings are reported in the following article. The results presented here are based on the data collected by Belle II during 2019-2021. }


\arxivnumber{1234.56789} % Only if you have one

\begin{document}
\maketitle
\flushbottom

\section{Introduction}
\label{sec:intro}
As the only leptons massive enough to decay into hadrons, taus not only allow to investigate the hadronization mechanism via their hadronic final states, but 
might preferentially couple to non-SM physics, through mass-dependent couplings. Therefore, any possible contribution from a new mediator whose coupling is proportional to lepton masses might be enhanced. From the experimental point of view though, taus are challenging since they can not be detected as long-lived particles, but must instead be reconstructed from their final-state products, which involve undetectable neutrinos.
Furthermore, they allow searching for charged lepton flavor violation (LFV), which would provide an indisputable proof for beyond SM physics. Processes involving LFV can occur in the SM via weak interaction charged currents, due to neutrino oscillations, and are predicted at the level of 10$^{-50}$, which is beyond the reach of current and future experiments. % Therefore any observation of LFV signatures would be an unambiguous hint of non-SM physics.
Belle II has a unique capability to probe both new invisible mediators and LFV in $\tau$ decays. Moreover, it can look for indirect signs of non-SM physics through high precision measurements of SM fundamental parameters. We report searches for new invisible particles, $\tau$ LFV decays and the measurement of the $\tau$ lepton mass using the data collected by the Belle II detector \cite{ref:tdr_belle} at the SuperKEKB asymmetric energy \epem\ collider. %~\cite{ref:superkekb}.
SuperKEKB mainly operates at a centre-of-mass energy (c.m.) of 10.58~GeV and adopts a nano-beam scheme to reach unprecedented instantaneous luminosity. % while almost doubling the beam currents, with the final goal to achieve an instantaneous luminosity of order $\sim 6\times10^{35}$cm$^{-2}$/s. %The final target is to deliver a 50 times larger data set with respect to what collected by Belle, for a total integrated luminosity of 50\invapto. 
At the time of this conference, the accelerator has achieved the peak luminosity world record of 4.7$\times 10^{34}$~cm$^{-2}/s$ and Belle II has so far collected 424\invfem\ of data, including unique energy scan samples. It is currently in its first long shutdown. 

\section{Leptons as discovery tools: the experimental challenges}
Leptonic production of tau pair processes \taupair\ provide a very clean physics environment and can rely on precise QED predictions to look for physics beyond the SM. The way is two-fold: one could look for deviations from SM predictions in high precision measurements of very clean and precisely computed observables; a second possibility is instead to search for processes that would be either forbidden or highly suppressed in the SM and whose observation is \textit{per se} a hint of new physics. The first class of measurements are mainly systematically limited and to improve the current results and attain the world's best precision, an excellent understanding of the experiment performance at the fraction of permille level is required. On the other hand, measurements of rare or forbidden processes imply to achieve unprecedented luminosity to collect sufficiently large data sets and devise new analysis techniques to boost signal efficiencies in order to reach sensitivities below the $10^{-8}$ level.

\section{Experimental facility: Belle II}\label{sec:exp}
The Belle II detector, the main upgrade of its predecessor Belle, is a multipurpose spectrometer  surrounding the \emep\ interaction point and providing coverage of more than 90\% of the solid angle. %It consists of a tracking system composed of a vertex detector, with two layers of pixels and four layers of double-sided silicon strips sensors, and a small-cells helium-based central drift chamber (CDC); a particle identification system and an electromagnetic calorimeter (ECL) for electron and photons reconstruction are also inside the the 1.5 T superconducting magnet. The outermost sub-system consists of a dedicated muon and $K^0_L$ detector (KLM). 
The details of the Belle II detector can be found elsewhere~\cite{ref:tdr_belle}. Belle II ensures a very high reconstruction efficiency for neutral particles and excellent resolutions despite the harsh beam background environment, both of which are crucial when dealing with recoiling system and missing-energy final states. Additionally, it is equipped with
dedicated low-multiplicty trigger lines at hardware level, mainly based on calorimetric information, which were not available at Belle. Profiting from the well known initial state of \epem\ collisions, and its near-hermetic coverage, Belle II has a unique capability to probe signatures involving invisible final states and particles escaping detection. 
 Moreover, the production cross-section for \taupair\ events is 0.919 nb at a c.m. energy \cmenergy, allowing Belle II to collect large data samples for precision measurements of $\tau$ lepton properties.
 
\subsection{Typical $\tau$ signatures in \emep\ collisions} 
In \taupair\ processes, tau candidates are produced back-to-back in c.m. system. Their decay products are well separated into two opposite hemispheres, defined by the plane perpendicular to the thrust axis $\mathbf{\hat{n}}_T$, which is the vector maximizing the quantity
\begin{equation}
\label{eq:thrust}
	T = \max_{\mathbf{\hat{n}}_T} \left(\dfrac{\sum_{i} \left|\mathbf{p}_i \cdot \mathbf{\hat{n}}_T\right|}{\sum_{i} \left|\mathbf{p}_i\right|} \right),
\end{equation}
where $\mathbf{p}_i$ is the momentum of the final state particle $i$, including both charged and neutral particles.
According to the number of charged particles in each hemisphere, consistently with charge conservation in $\tau$ decays, two main topologies can be selected: the $3\times1$-prong decays, with three charged particles on one side and only one in the opposite hemisphere; or the $1\times1$-prong decays.
\begin{figure}[htbp]
\centering
\includegraphics[width=.44\textwidth]{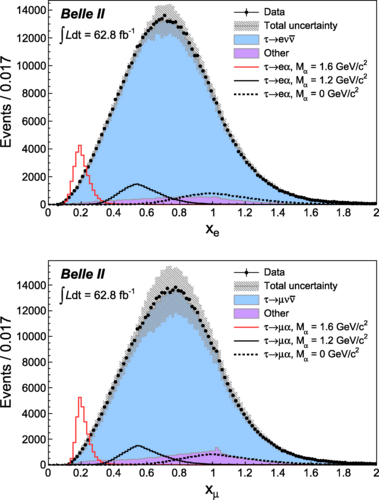}\qquad
\includegraphics[width=.47\textwidth]{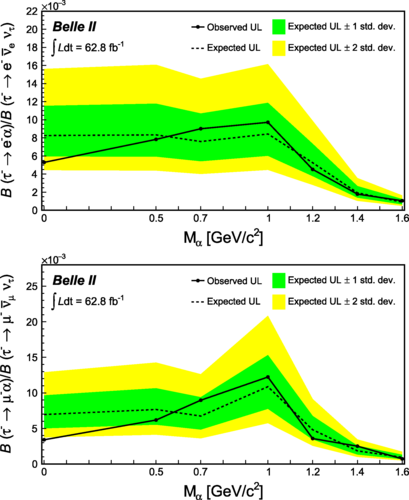}
\caption{On the left, the distribution of the normalized lepton energy $x_{\ell}$ for the electron (top) and muon (bottom) channel in the search for $\tau\to \ell\alpha$ is shown. Data are the black dots and the simulation is the stack filled histograms. On the right, upper limits at 95\% C.L. on the branching-fraction ratios $\mathcal{B}(\tau \rightarrow e \alpha) /\mathcal{B}(\tau \to e \nu \bar{\nu})$ (top) and $\mathcal{B}(\tau \to \mu \alpha)/\mathcal{B}(\tau \to \mu \nu \bar{\nu})$ (bottom) as a function of the $\alpha$ mass, as well as their expectations from background-only hypothesis. }\label{fig:topology}
\end{figure}
Requiring the reconstructed tracks to match one of these topology classes is a powerful way to suppress the main background from continuum \qqbar\ processes and enhance signal purity when reconstructing \taupair\ events.

\section{Searches for a new invisible boson $\alpha$ in $\tau$ decays }
%\subsection{Searches for lepton flavor violation in $\tau$ decays}
Decays of $\tau$ leptons to new LFV bosons are postulated in many models~\cite{ref:invboson}. The process searched for in this study is \taulalpha\ and its charge conjugated. The signal $\tau$ is searched for in its decay to a new invisible boson $\alpha$, accompanied by a lepton $\ell = e, \mu$, therefore $3\times1$-prong events are selected. The signal $\tau$ rest-frame is approximated using as energy half the collision energy $\sqrt{s}/2$ and as momentum direction the opposite to the one of the reconstructed tag $\tau$. We exploit the kinematic features of the signal process as a two-body decay to discriminate it from the background, by looking for a narrow peak in the distribution of the normalized lepton energy in the c.m. frame (Figure~\ref{fig:topology}, left plot) over a smooth contribution coming from the irreducible background of $\tau\to \ell \nu \bar{\nu_{\ell}}$ processes. In absence of any signal excess in 63\invfem data, 95\% CL upper limits in the mass range between 0 and 1.6 \gevcc\ are computed on the ratio of branching fractions $\mathcal{B}(\tau \to \ell \alpha)$ normalized to  $\mathcal{B}(\tau \to\ell \nu \bar{\nu_{\ell}})$~\cite{ref:taulalpha}. This analysis provides limits (Figure~\ref{fig:topology}, right plots) between 2-14 times more stringent than the previous one set by ARGUS~\cite{ref:argus}. 
% lphi
\section{Direct searches for LFV $\tau \to \ell\phi$ decays }
Possible new mediators may enhance the branching fraction for $\tau$ LFV decays \lphi\ up to observable levels of $\sim 10^{-8}$, and accommodate for flavor anomalies observed in lepton flavor universality tests with $B$ decays~\cite{ref:vlq_lphi}. In contrast to previous searches for \lphi\ decays performed at Belle~\cite{ref:belle_lphi} on \taupair\ events, we apply for the first time an \textit{untagged} approach. Only the signal $\tau$ decay to a $\phi$ meson candidate and a lepton, either muon  or electron, is explicitly reconstructed and the other $\tau$ is not required to decay to any specific known final state. Event kinematics features and signal properties are used in a BDT classifier to suppress the background, with twice the  signal efficiency for the muon mode with respect to previous analyses. Yields are extracted with a Poisson counting experiment approach from windows peaking at the known $\tau$ mass and at zero in the 2D plane of $(M_{\tau}, \Delta E_{\tau})$, respectively, where $\Delta E_{\tau}$ is the difference between the reconstructed energy of the signal $\tau$ in the c.m.~frame and half the collision energy. %The final yields selected in the data (black dots) and  in the simulation (light blue filled circles) after the box opening are shown in the left and right plots of Figure~\ref{fig:lphiyields} for the electron and muon channels, respectively. 
We find no significant excess and set 90\% CL upper limits on the branching fractions of $\mathcal{B}_{\mathrm{UL}}(\tau \to e\phi)=23\times 10^{-8}$ and $\mathcal{B}_{\mathrm{UL}}(\tau \to \mu \phi)=9.7\times 10^{-8}$~\cite{ref:taulphi}. 

\begin{comment} 
\begin{figure}[htbp]
\centering
\includegraphics[width=.45\textwidth]{images/unblindElectron.png}
\qquad
\includegraphics[width=.45\textwidth]{images/unblindMuon.png}
\caption{Scatter plots of \deltae\ vs \invmass\ for simulated SM background and data in the \myprocess\ (left) and  (right) channels, after all selections. The red squares represent the signal regions, while the green lines delineate the sidebands used for estimating the expected number of events in the SR.\label{fig:lphiyields}}
\end{figure}

\end{comment}

\section{Measurement of the $\tau$ lepton mass}
\begin{figure}[htbp]
\centering
\includegraphics[width=.45\textwidth]{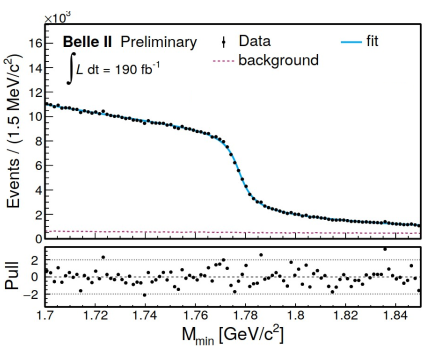}
\qquad
\includegraphics[width=.45\textwidth]{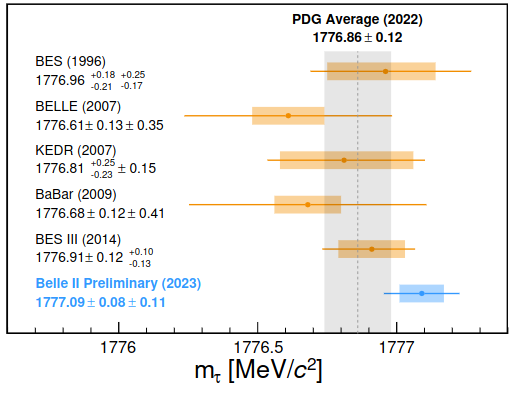}
\caption{On the left, the spectrum of the reconstructed pseudomass in data (black dots) and the superimposed fit (solid blue line) are shown. The bottom inset plot displays differences between data
and fit result divided by the statistical uncertainties. On the right, the most precise measurements of the tau mass to date, compared to the world average (gray band) and this work result (blue text). \label{fig:tauMass}}
\end{figure}

%tau mass
 Lepton properties are fundamental parameters of the SM and need to be measured with the highest precision. By applying the pseudo-mass $M_{min}$ technique to reconstructed \taupair\ events from 190\invfem\ data, we provide the world's most precise measurement of the $\tau$ mass $M_{\tau}$. The measured value is extracted from a fit to the endpoint of the distribution $M_{min}=\sqrt{M_{3\pi}^2+2(\sqrt{s}/2-E^*_{3\pi})(E^*_{3\pi}-P^*_{3\pi})}$, which is computed from events where the signal $\tau$ is reconstructed in its decays to three charged pions and the other $\tau$ decaying into one charged particle. The distributions of the pseudomass in simulation and data is shown in the left plot of Figure~\ref{fig:tauMass}. An excellent control of the systematic sources, dominated by the calibration of the beam energies and the charged-particle momentum scale, is required to reduce the total systematic uncertainty to 0.11~\mevcc, achieving the most precise measurement to date of the $\tau$ lepton mass of $1777.09\pm 0.08_{\mathrm{stat}}\pm 0.11_{\mathrm{sys}}$~\cite{ref:taumass}.
%lfu test
\section{Prospects on lepton flavor universality tests}
Lepton flavor universality (LFU) in SM assumes all three leptons have equal coupling strength to the charged gauge bosons of the electroweak interaction. Many models predict new forces violating LFU%, that could for example imply a new singly-charged scalar singlet
~\cite{ref:lfuScalar}. Tau decays allow high precision tests of LFU by measuring the ratio of the branching fractions of $\tau$ decays to muon and to electron,  
\begin{equation}
\centering
R_{\mu} =\frac{\mathcal{B}(\tau\to\mu\nu_{\mu}\nu_{\tau})}{\mathcal{B}(\tau\to e\nu_{e}\nu_{\tau})}
\end{equation}
The most precise result to date is $R_{\mu} = 0.9796 \pm 0.0016_{\mathrm{stat}} \pm 0.0036_{\mathrm{sys}}$ provided by Babar~\cite{ref:lfuBabar}. It uses 467\invfem\ collision data,
%to measure $R_{\mu}$ from $3\times1$-prong decays in \taupair\ events 
for a final 0.4\% precision, systematically dominated by the contribution of the particle identification (PID) and trigger uncertainties. Simulation studies at Belle II show room for several improvements: by devising dedicated low multiplicity triggers based on calorimeter information, which provide a better understanding of the kinematic dependency and reduce the associated systematic uncertainty; by dropping the likelihood-based PID selector for pions and deploying BDT classifier for lepton identification, which will decrease the probability to wrongly identify a pion as a lepton to less than 0.1\%; eventually, adding the $1\times1$-prong decays as signal signature to increase the size of the analyzed data set. Further studies for the development of the specific 1x1 topology triggers are still needed, but already with one quarter of Babar data set, Belle II expected sensitivity achieves the same statistical precision of 0.16\%.
% Bibliography

%% [A] Recommended: using JHEP.bst file
\bibliographystyle{JHEP}
\bibliography{biblio.bib}

%% or
%% [B] Manual formatting (see below)
%% (i) We suggest to always provide author, title and journal data or doi:
%% in short all the informations that clearly identify a document.
%% (ii) please avoid comments such as "For a review'', "For some examples",
%% "and references therein" or move them in the text. In general, please leave only references in the bibliography and move all
%% accessory text in footnotes.
%% (iii) Also, please have only one work for each \bibitem.

\end{document}